\documentclass[apj,twocolumn,twocolappendix,numberedappendix]{openjournal}
\usepackage{amsmath}
\usepackage{booktabs}
\usepackage{multirow}
\usepackage{color}
\usepackage{soul}
\usepackage{booktabs} 
\usepackage{threeparttable}

\usepackage{siunitx}
\usepackage{enumitem}

\usepackage[dvipsnames]{xcolor} 

\usepackage[breaklinks,colorlinks,citecolor=blue,urlcolor=blue]{hyperref}
\usepackage{xurl}

\renewcommand{\arraystretch}{1.2}
\usepackage{listings}
\usepackage{color}
\definecolor{dkgreen}{rgb}{0,0.6,0}
\definecolor{gray}{rgb}{0.5,0.5,0.5}
\definecolor{mauve}{rgb}{0.58,0,0.82}
\definecolor{golden}{rgb}{0.86,0.65,0.01}
\usepackage{orcidlink}

\begin{document}


\title{\textbf{ASAS-SN R\lowercase{ates} IV: C\lowercase{onstraints on the} K\lowercase{ilonova} R\lowercase{ate}}}

\author{\vspace{-1.4cm}
        D.~D.~Desai\,\orcidlink{0000-0002-2164-859X}$^{1}$}
\author{B.~J.~Shappee\,\orcidlink{0000-0003-4631-1149}$^{1}$}
\author{C.~S.~Kochanek\,\orcidlink{}$^{2,3}$}
\author{K.~Z.~Stanek$^{2,3}$}
\author{K.~Auchettl\,\orcidlink{0000-0002-4449-9152}$^{4,5}$}
\author{J.~F.~Beacom\,\orcidlink{0000-0002-0005-2631}$^{6,2,3}$}
\author{J.~Cooke$^{7,8,9}$}
\author{Subo~Dong\,\orcidlink{0000-0002-1027-0990}$^{10,11,12}$}
\author{W.~B.~Hoogendam\,\orcidlink{0000-0003-3953-9532}$^{1}$}
\author{J.~L.~Prieto$^{13,14}$}
\author{T.~A.~Thompson\,\orcidlink{0000-0003-2377-9574}$^{2,6,3}$}
\author{M.~A.~Tucker\,\orcidlink{0000-0002-2471-8442}$^{3,2,6}$}
\author{N.~Van~Bemmel\,\orcidlink{0000-0002-2194-4567}$^{7,8}$ \vspace{0.15cm}}

\affiliation{$^1$ Institute for Astronomy, University of Hawai‘i, Honolulu, HI, USA}

\affiliation{$^{2}$ Department of Astronomy, The Ohio State University, Columbus, OH, USA}

\affiliation{$^{3}$ Center for Cosmology and AstroParticle Physics, The Ohio State University, Columbus, OH, USA}

\affiliation{$^{4}$ OzGrav, School of Physics, The University of Melbourne, Parkville, Victoria, Australia}

\affiliation{$^{5}$ Department of Astronomy and Astrophysics, University of California, Santa Cruz, CA, USA}

\affiliation{$^{6}$ Department of Physics, The Ohio State University, Columbus, USA}

\affiliation{$^{7}$ Centre for Astrophysics and Supercomputing, Swinburne University of Technology, Victoria, Australia}

\affiliation{$^{8}$ ARC Centre of Excellence for Gravitational Wave Discovery (OzGrav), Victoria, Australia}

\affiliation{$^{9}$ Australian Research Council Centre of Excellence for All-sky Astrophysics in 3 Dimensions (ASTRO 3D), Australia}

\affiliation{$^{10}$ Department of Astronomy, School of Physics, Peking University, Beijing, China}

\affiliation{$^{11}$ Kavli Institute of Astronomy and Astrophysics, Peking University, Beijing, China}

\affiliation{$^{12}$ National Astronomical Observatories, Chinese Academy of Science, Beijing, China}

\affiliation{$^{13}$ Instituto de Estudios Astrof\'isicos, Facultad de Ingenier\'ia y Ciencias, Universidad Diego Portales, Santiago, Chile}

\affiliation{$^{14}$ Millennium Institute of Astrophysics MAS, Santiago, Chile}

\email{Corresponding author: dddesai@hawaii.edu}

\begin{abstract}
Kilonovae (KNe) are the electromagnetic signatures of neutron star mergers and are likely the dominant site of cosmic $r$-process nucleosynthesis. However, their intrinsic rate remains poorly constrained due to a paucity of confirmed events. We use the All-Sky Automated Survey for Supernovae (ASAS-SN) to place limits on the rate of bright, nearby KNe over an 11-year baseline ranging from 2014 to 2024. To evaluate the survey's completeness for KNe, we employ an injection-recovery simulation using a shock-cooling cocoon model calibrated to the early blue emission of the only well-sampled KN, SSS17a (AT~2017gfo). Finding no KNe within the survey, we calculate a $2\sigma$ ($\sim95\%$) upper limit on the local volumetric KN rate of $R_{\mathrm{KN}} < 4400\,\mathrm{yr}^{-1}\,\mathrm{Gpc}^{-3}$. Despite ASAS-SN's shallower limiting magnitude compared to other time-domain searches, its continuous, high-cadence, all-sky monitoring yields a constraint that is competitive with the strongest results from electromagnetic surveys but remains a factor of 18 higher than the LIGO-Virgo-KAGRA GWTC-4 estimate of the binary neutron star merger rate.


\end{abstract}

\maketitle

\section{Introduction} \label{sec:intro}

Kilonovae (KNe) are electromagnetic transients powered by the merger of a binary neutron star (BNS) or a neutron star (NS) and a black hole \citep[BH;][]{li_transient_1998,metzger_electromagnetic_2010,roberts_electromagnetic_2011,barnes_effect_2013}. They are likely the dominant producers of the heaviest elements in the Universe \citep[e.g.,][]{arnould_r-process_2007,johnson_populating_2019}. Observations of the gravitational wave (GW) event GW170817 \citep{abbott_gw170817_2017} found an optical/infrared counterpart, SSS17a (AT~2017gfo; \citealp{coulter_swope_2017}), consistent with theoretical predictions for the electromagnetic emission from a binary NS merger \citep[e.g.,][]{abbott_multi-messenger_2017,arcavi_optical_2017,chornock_electromagnetic_2017,coulter_swope_2017,cowperthwaite_electromagnetic_2017,drout_light_2017,kasen_origin_2017,kasliwal_illuminating_2017,kilpatrick_electromagnetic_2017,murguia-berthier_neutron_2017,mccully_rapid_2017,nicholl_electromagnetic_2017,pan_old_2017,shappee_early_2017,smartt_kilonova_2017,soares-santos_electromagnetic_2017,tanaka_kilonova_2017,tanvir_emergence_2017,piro_evidence_2018}. This remains the only electromagnetically and spectroscopically confirmed KN associated with a GW event.

Aside from SSS17a (AT~2017gfo), other KN candidates detected from the afterglows of short and long gamma-ray bursts (GRBs) are typically too distant and faint to obtain well-sampled light curves in the optical and infrared \citep[e.g.,][]{tanvir_kilonova_2013, berger_r-process_2013, yang_possible_2015, jin_macronova_2016, troja_luminous_2018, lamb_short_2019, rastinejad_kilonova_2022, mei_gigaelectronvolt_2022, levan_long-duration_2023, sun_magnetar_2025, nugent_where_2025}. Consequently, the intrinsic rate of KNe remains poorly constrained and the estimates derived from short GRBs and GW detections still span more than an order of magnitude \citep[e.g.,][]{fong_decade_2015, dellavalle_gw170817_2018, zhang_peculiar_2018, dichiara_short_2020, abbott_gwtc-2_2021, mandel_rates_2022,abbott_population_2023, abac_gwtc-40_2025}. 

While follow-up campaigns of GW triggers are efficient \citep[including the search by ASAS-SN;][]{dejaeger_asas-sn_2022}, they are limited by merger orientation, the size and visibility of the localization regions, and the operational status of the GW detectors. Independent optical surveys offer a complementary approach by scanning large areas of the sky without relying on external triggers \citep[e.g.,][]{perkins_searching_2026}. Previous searches by surveys such as the Asteroid Terrestrial-impact Last Alert System (ATLAS), the Dark Energy Survey (DES), the Palomar Transient Factory (PTF), and the Zwicky Transient Facility (ZTF) have not yet yielded a confirmed KN detection, constraining the rate to roughly $< 800 - 99{,}000\,\text{yr}^{-1}\,\text{Gpc}^{-3}$ depending on the survey depth and area \citep{doctor_search_2017, kasliwal_illuminating_2017, smartt_kilonova_2017, andreoni_fast-transient_2021}. Most recently, the Kilonova and Transients Programme (KNTraP) used the Dark Energy Camera (DECam) to search for KNe out to $z \sim 0.3$, placing an upper limit of $< 180{,}000\,\text{yr}^{-1}\,\text{Gpc}^{-3}$ from their first observing run \citep{vanbemmel_optically_2025}.

While the All-Sky Automated Survey for Supernovae \citep[ASAS-SN;][]{shappee_man_2014, kochanek_all-sky_2017, hart_asas-sn_2023} has become known for its discoveries of interesting events --- such as tidal disruption events \citep[e.g.,][]{holoien_six_2016,holoien_asassn-15oi_2016,holoien_unusual_2018,holoien_discovery_2019,holoien_rise_2020,holoien_investigating_2022,hinkle_discovery_2021,hinkle_curious_2022,payne_asassn-14ko_2021,payne_rapid_2022,hoogendam_discovery_2024}, superluminous supernovae \citep{dong_asassn-15lh_2016}, nearby supernovae \citep[SNe; e.g.,][]{shappee_young_2016, shappee_seeing_2018, vallely_asassn-18tb_2019}, novae and dwarf novae \citep[e.g.,][]{kato_survey_2014-1, kato_survey_2014, kato_survey_2015, kato_survey_2016, kato_survey_2017, li_nova_2017, aydi_direct_2020, aydi_early_2020, kawash_classical_2021}, large M-dwarf flares \citep[e.g.,][]{schmidt_characterizing_2014,schmidt_asassn-16ae_2016,schmidt_largest_2019,rodriguez_martinez_catalog_2020}, and variable stars \citep[e.g.,][]{jayasinghe_asas-sn_2018,jayasinghe_asas-sn_2019,jayasinghe_asas-sn_2019-1,jayasinghe_asas-sn_2020-2,jayasinghe_asas-sn_2020,jayasinghe_asas-sn_2020-1,jayasinghe_asas-sn_2021,christy_asas-sn_2023} --- it was designed to accurately measure local transient rates. ASAS-SN occupies a unique niche in the landscape when compared to the KN surveys listed above. Although limited to a shallower magnitude ($g \sim 18$ mag) than deeper surveys like ZTF or KNTraP, ASAS-SN provides continuous, high-cadence monitoring of the entire visible sky over a very long baseline, which is ideal for rate measurements of transients \citep[e.g.,][]{brown_relative_2019, kawash_galactic_2022, desai_supernova_2024, pessi_supernova_2025, desai_supernova_2026}.

In this work, we constrain the local KN rate using ASAS-SN data spanning 2014--2024. Finding no KNe, we calculate an upper limit on the local volumetric KN rate. In Section~\ref{sec:LC}, we describe the KN light curve templates used for our simulations. In Section~\ref{sec:rates}, we detail our completeness analysis and present the resulting rate upper limit. We discuss our results in Section~\ref{sec:summary}. Throughout this analysis, we adopt a flat $\Lambda$CDM cosmology with $H_0 = 70~\mathrm{km~s^{-1}~Mpc^{-1}}$ and $\Omega_{m,0} = 0.3$.

\section{Light Curve Template} \label{sec:LC}

The optical and infrared evolution of a KN is largely governed by the radioactive $\beta$-decay of $r$-process elements synthesized in the neutron-rich ejecta \citep{arnould_r-process_2007}. Their appearance is heavily dependent on the composition, opacity, and energy deposition of the ejected material \citep[e.g.,][]{metzger_electromagnetic_2010, barnes_effect_2013, barnes_radioactivity_2016, kasen_radioactive_2019, barnes_kilonovae_2021, desai_plasma_2025}. Theoretical models generally predict two distinct components: a ``red'' component and a ``blue'' component. The red component, associated with high opacity lanthanide-rich dynamical ejecta, evolves slowly over days to weeks and peaks in the infrared. The ``blue'' component arises from lanthanide-poor material, such as accretion disk winds or shock-heated ejecta, and is characterized by lower opacities, rapid evolution, and a peak in the UV/optical bands on timescales of hours to days \citep[e.g.,][]{metzger_electromagnetic_2010, kasen_opacities_2013, tanaka_radiative_2013, grossman_long-term_2014, piro_evidence_2018}.

\begin{figure}
    \centering
    \includegraphics[width=\columnwidth]{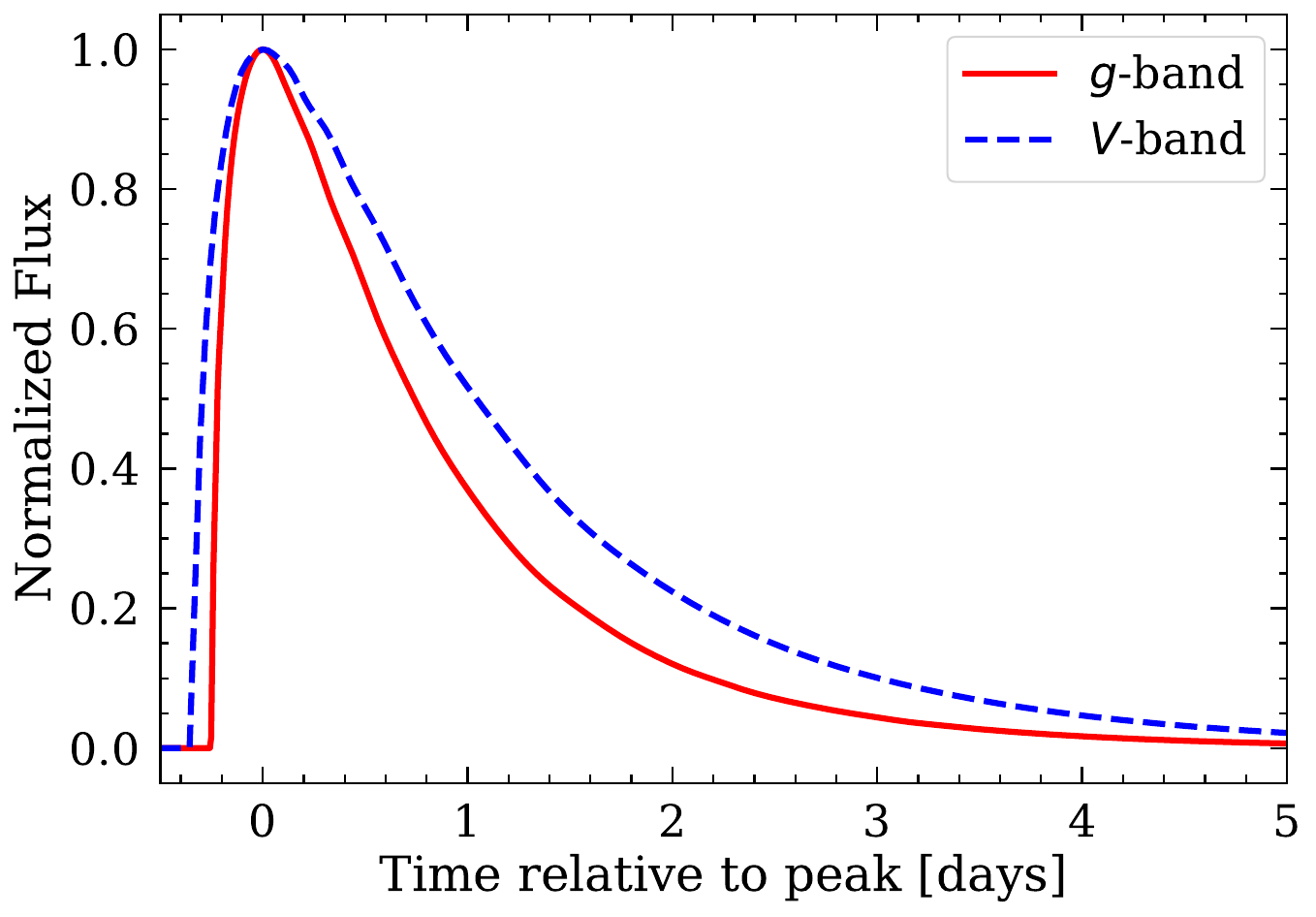}
    \caption{Normalized $V$- and $g$-band KN light curve templates from the shock-cooling cocoon model of \citet{piro_evidence_2018}, which explain the early, blue emission from SSS17a (AT~2017gfo). The templates are shifted such that the peak is at $t=0$~days.}
    \label{fig:LC_template}
\end{figure}

Given ASAS-SN's use of the relatively blue optical $V$ and $g$ bands and a current limiting magnitude of $g \sim 18$~mag, the survey is primarily sensitive to the bright, peak optical emission of the blue component. Therefore, to model the detectability of KNe in our data, we use the shock-cooling cocoon model of \citet{piro_evidence_2018}, which was developed to explain the early, blue emission of the best characterized and the only unambiguously confirmed KN associated with a GW event, SSS17a (AT~2017gfo). Figure~\ref{fig:LC_template} shows these $V$- and $g$-band light curves. We review other possible models in Section~\ref{sec:summary}.

Physically, this model attributes the early luminosity to the cooling of material shock-heated by the interaction of the gamma-ray burst jet with the merger debris, the so-called ``cocoon'' \citep{piro_evidence_2018}. As the jet punches through the ejecta, it deposits energy, creating a wide-angle outflow. \citet{piro_evidence_2018} demonstrate that the rapid rise and power-law decay of the luminosity, temperature, and photospheric radius observed in SSS17a (AT~2017gfo) during the first $\sim 3.5$ days are best described by the cooling of this shocked layer, rather than pure radioactive heating.

\section{Rate Constraints} \label{sec:rates}

To characterize the ASAS-SN discovery and independent recovery efficiency for KNe, we adopt the methodology established in our previous SNe rate studies \citep{desai_supernova_2024,pessi_supernova_2025, desai_supernova_2026} and summarize it in this section. We employ an injection-recovery simulation to determine the survey completeness for KN light curves as a function of time and apparent magnitude.

\begin{figure}
    \centering
    \includegraphics[width=\columnwidth]{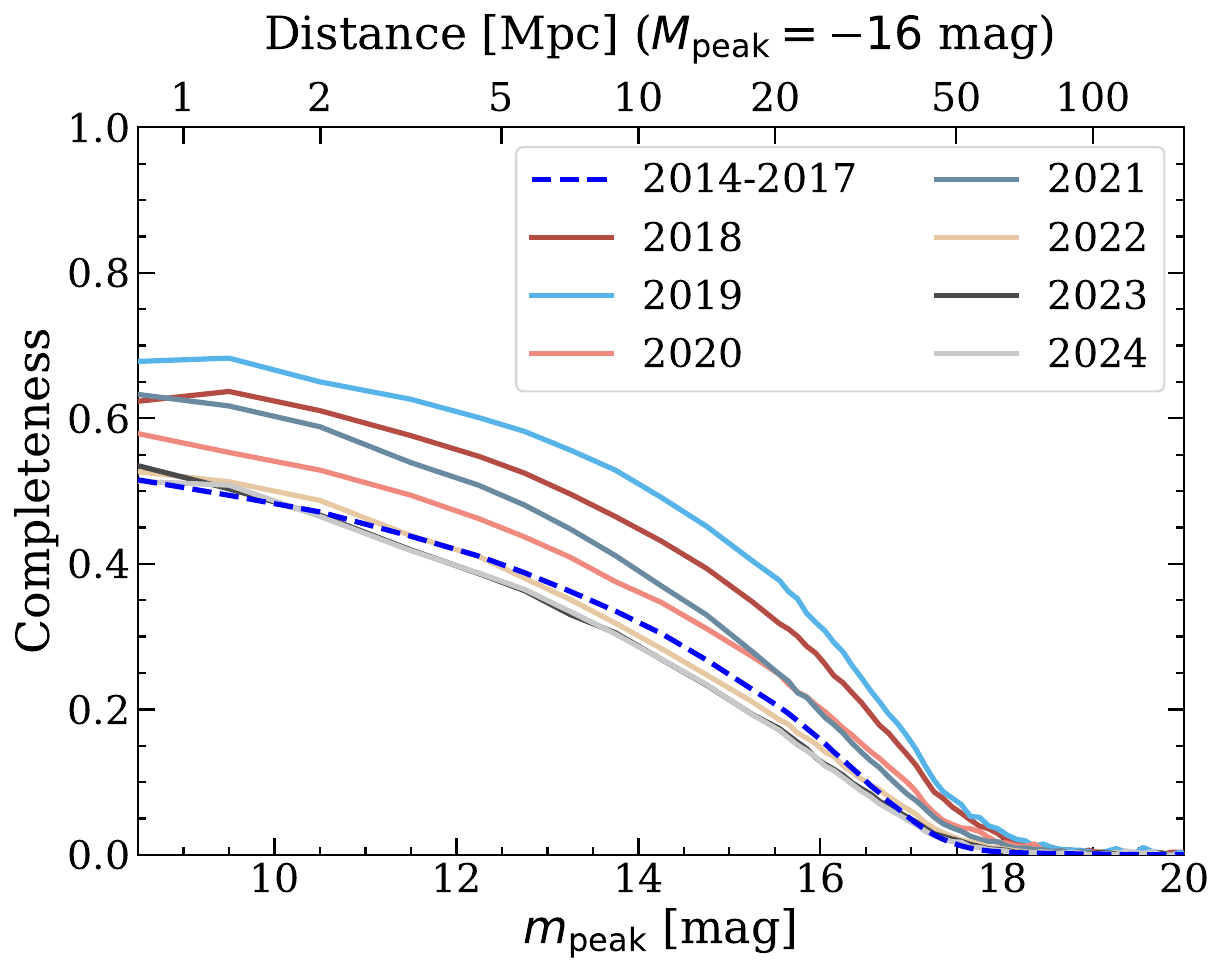}
    \caption{Completeness as a function of time and peak apparent magnitude for KNe. The dashed curve shows the $V$-band era (2014--2017), while the solid curves show individual years of the $g$-band era (2018--2024). The top axis indicates the distance corresponding to an event with a peak absolute magnitude of $M_{\mathrm{peak}}=-16$~mag, similar to SSS17a (AT~2017gfo).}
    \label{fig:completeness}
\end{figure}

We first extract ASAS-SN light curves at $10^5$ random coordinates distributed uniformly across the entire sky. These baseline light curves do not trace specific astrophysical sources; instead, they empirically capture the true observing conditions of the survey, including the photometric noise, observing cadence, weather interruptions, and seasonal gaps. We inject the simulated $V$- and $g$-band KN templates described in Section~\ref{sec:LC} into these baseline light curves. For each of the $10^5$ sky locations, we perform 200 random trials (randomly selected in the comoving volume and time of peak) by adding the simulated transient flux to the measured background flux.

A trial's `recovery' is not defined by a simple photometric threshold (e.g., $>5\sigma$). Instead, we calculate the recovery probability threshold for each epoch using the empirically derived efficiency curves where ASAS-SN could discover or independently recover a transient. The ASAS-SN baseline spans two distinct phases: the $V$-band era (2014--late-2017) and the $g$-band era (late-2017--2024). Separately for each era, we model the signal-to-noise ratio (SNR)-dependent single-epoch discovery/recovery efficiency, $p(\mathrm{SNR})$. For the $V$-band data, we employ the piecewise function derived in \citet{desai_supernova_2024} peaking at $p=0.65$, while for the $g$-band data, we use the yearly skew-normal efficiency models from \citet{desai_supernova_2026}, peaking at $p=0.48$ averaged across all years. The latter explicitly accounts for variations in survey depth and operational changes over time, such as those occurring during the COVID-19 pandemic.

By modeling the discovery/recovery efficiency as a continuous function of SNR, this method accounts for pipeline systematics and vetting efficiency, capturing the probabilistic nature of transient discovery/recovery often missed by idealized threshold cuts. An injected KN trial is considered recovered if at least one epoch meets the discovery/recovery criteria described above. Figure~\ref{fig:completeness} shows the resulting completeness (the fraction of total trials that are recovered) as a function of peak apparent magnitude. The $g$-band years generally achieve higher completeness than the $V$-band years, primarily due to the higher cadence ($\sim1$--$2$~days) compared to the $V$-band era ($\sim2$--$3$~days) and a deeper limiting magnitude. While the $V$-band KN template is slightly broader than the $g$-band template, this morphological difference does not significantly impact the relative completeness because the rapid evolution of the transient is comparable to the cadence in both filters.

We systematically re-evaluated all unclassified transients flagged by the ASAS-SN pipeline to ensure that no KNe brighter than $g\sim18$~mag were missed or misclassified. We find no viable KN candidates. A detailed description of this verification process is provided in Appendix~\ref{app:unclass_verify}.

Since no KNe were detected in the 11-year baseline, we calculate an upper limit on the volumetric rate. We perform a Monte Carlo integration to determine the total effective volume-time ($VT$) probed by the survey. For this, we simulate $N_{\mathrm{sim}} = 10^6$ sources uniformly distributed in comoving volume within the redshift range $0.001 < z < 0.08$ and with peak absolute magnitudes drawn uniformly from the bin $M_{\mathrm{peak}} \in [-16.5, -15.5]$~mag, centered on the approximate luminosity of SSS17a (AT~2017gfo). We restrict the simulation to Galactic latitudes $|b| > 15^{\circ}$ to avoid regions of high extinction. We account for each simulated event by summing the completeness contributions from all available epochs to obtain a total effective volume-time
\begin{equation} \label{eq:vol_time}
VT = (1-\sin{b_\mathrm{lim}}) \frac{V_{\mathrm{sim}}}{N_{\mathrm{sim}}} \sum_{i=1}^{N_{\mathrm{sim}}} \left( \sum_{j} F_{i,j} \Delta t_j \right),
\end{equation}
where $(1-\sin{b_\mathrm{lim}})$ is the sky fraction correction for the Galactic plane cut, $V_{\mathrm{sim}}$ is the total simulation volume, $F_{i,j}$ is the completeness of the $i^{\mathrm{th}}$ source in the $j^{\mathrm{th}}$ epoch, and $\Delta t_j$ is the duration of that epoch. In total, we divide the 11-year baseline into eight distinct epochs: a single epoch for the $V$-band era (2014--2017) and seven individual yearly epochs for the $g$-band era (2018--2024).

We define our upper limit as the volumetric rate, $R_{\mathrm{KN}}$, corresponding to a Poisson expectation of the number of KNe, $N_{\mathrm{exp}}$, over the duration of the survey. Based on the total effective volume-time calculated from Equation~\ref{eq:vol_time}, we find a $2\sigma$ ($N_{\mathrm{exp}}=3.08$) upper limit on the local KN rate of
\begin{equation} \label{eq:rate_limit}
R_{\mathrm{KN}} < 4400\,\mathrm{yr}^{-1}\,\mathrm{Gpc}^{-3}
\end{equation}
for ``blue'' SSS17a-like events. This constraint is presented in Figure~\ref{fig:rate} alongside the $95\%$ confidence limits from other optical surveys. The $1\sigma$ ($N_{\mathrm{exp}}=1.15$) and $3\sigma$ ($N_{\mathrm{exp}}=5.81$) upper limits are $R_{\mathrm{KN}} < 1700\,\mathrm{yr}^{-1}\,\mathrm{Gpc}^{-3}$ and $R_{\mathrm{KN}} < 8500\,\mathrm{yr}^{-1}\,\mathrm{Gpc}^{-3}$, respectively. 

To provide an intuitive context, we express the rate limit in terms of Milky Way-like galaxies. Assuming a density scale for $L_*$ galaxies of $n_* \approx 0.01\,h^3\,\mathrm{Mpc^{-3}}$ \citep[e.g.,][]{blanton_galaxy_2003, bell_optical_2003} and adopting $h=0.7$, our $2\sigma$ volumetric upper limit translates to a rate per $L_*$ galaxy of $R_{\mathrm{KN}}n_*^{-1} < 0.13 \,\mathrm{century^{-1}}\,(L_*\,\mathrm{galaxy)^{-1}}$. This alternative scaling is also shown on the right axis of Figure~\ref{fig:rate}.

We also place this KN rate limit in the broader context of supernova rates measured by ASAS-SN: $R_{\mathrm{Ia}} = (2.55 \pm 0.12) \times 10^4\,\mathrm{yr}^{-1}\,\mathrm{Gpc}^{-3}$ for Type Ia supernovae \citep{desai_supernova_2026} and $R_{\mathrm{CC}} = 7.0_{-0.9}^{+1.0} \times 10^4\,\mathrm{yr}^{-1}\,\mathrm{Gpc}^{-3}$ for core-collapse supernovae \citep{pessi_supernova_2025}. Comparing these values, our $2\sigma$ upper limit constrains the KN rate to be lower than the Type Ia and core-collapse rates by factors of at least 6 and 16, respectively.

\section{Discussion} \label{sec:summary}

\begin{figure*}
    \centering
    \includegraphics[width=\textwidth]{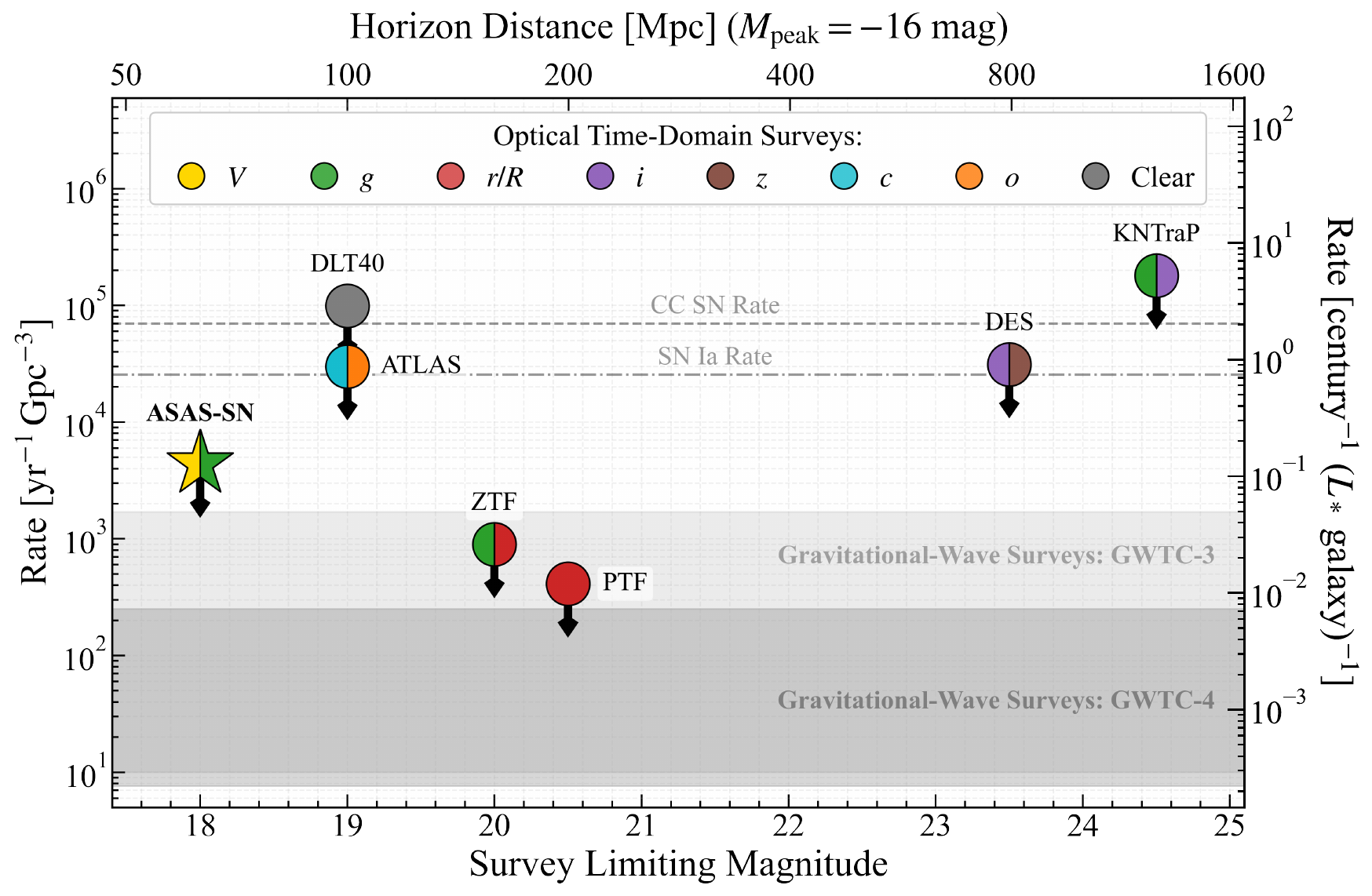}
    \caption{Comparison of the ASAS-SN $95\%$ volumetric KN rate upper limit ($R_{\mathrm{KN}} < 4400\,\mathrm{yr}^{-1}\,\mathrm{Gpc}^{-3}$) with the $95\%$ confidence limits from other optical surveys: ATLAS \citep{smartt_kilonova_2017}, PTF \citep{kasliwal_illuminating_2017}, DES \citep{doctor_search_2017}, DLT40 \citep{yang_empirical_2017}, ZTF \citep{andreoni_fast-transient_2021}, and KNTraP \citep{vanbemmel_optically_2025}. The color of the point indicates the filters used for the KNe search in those surveys. The top axis illustrates the corresponding horizon distance for an SSS17a-like event with a peak absolute magnitude of $M_{\mathrm{peak}} = -16$~mag. The light and dark shaded bands show the estimate of the local ($z=0$) BNS merger rate from the LIGO-Virgo-KAGRA collaboration GWTC-3 \citep[$10 - 1700\,\mathrm{yr}^{-1}\,\mathrm{Gpc}^{-3}$;][]{abbott_population_2023} and GWTC-4 \citep[$7.6 - 250\,\mathrm{yr}^{-1}\,\mathrm{Gpc}^{-3}$;][]{abac_gwtc-40_2025}, respectively. The dashed and dash-dotted lines show the ASAS-SN core-collapse and Type Ia SN rates, respectively \citep[][]{pessi_supernova_2025,desai_supernova_2026}. The KN rate inferred from short GRB observations ranges between $160 - 352\,\mathrm{yr}^{-1}\,\mathrm{Gpc}^{-3}$ with $>60\%$ fractional uncertainties \citep{fong_decade_2015, dellavalle_gw170817_2018, zhang_peculiar_2018, dichiara_short_2020}.}
    \label{fig:rate}
\end{figure*}

With the lack of detected KNe in the 11-year ASAS-SN baseline, we place a $2\sigma$ ($\sim95\%$) upper limit on the local volumetric KN rate of $R_{\mathrm{KN}} < 4400\,\mathrm{yr}^{-1}\,\mathrm{Gpc}^{-3}$. While ASAS-SN's limiting magnitude ($g\sim 18$~mag) is shallower than that of other time-domain searches, our constraint is highly competitive, falling within an order of magnitude of limits derived from other surveys. ASAS-SN benefits from a very long baseline coupled with high-cadence, all-sky observations. Furthermore, this analysis benefits from a detailed characterization of the systematics affecting completeness. Moreover, ASAS-SN was nearly spectroscopically complete for the $V$-band era \citep{holoien_asas-sn_2017-2,holoien_asas-sn_2017-1,holoien_asas-sn_2017,holoien_asas-sn_2019} and $\sim81\%$ complete on average for the $g$-band era \citep{neumann_asas-sn_2023,desai_supernova_2026}.

Figure~\ref{fig:rate} illustrates the $2\sigma$ ($\sim95\%$) ASAS-SN limit in the context of other independent, optical searches. The top axis denotes the horizon distance --- the maximum distance at which an SSS17a-like event with a peak absolute magnitude of $M_{\mathrm{peak}} = -16$~mag could be detected. For ASAS-SN, this horizon is approximately $60$~Mpc ($\sim20$~Mpc further than SSS17a); at this distance, any detected events would be sufficiently bright to be characterized in great detail. While our analysis requires a transient to be recovered in only a single epoch, this requirement is strictly defined by an empirically derived recovery efficiency rather than a simple signal-to-noise threshold. ATLAS and DLT40 require a single detection, whereas DES requires paired detections in the $i$ and $z$ bands on the same or adjacent nights. ZTF requires at least two detections to confirm a rapid decline, while KNTraP and PTF require three detections to reject cosmic rays and image artefacts. The upper limits from KNTraP, ATLAS, DLT40, and ZTF are for $95\%$ confidence, while the DES and PTF reported limits are for $90\%$ and $99.7\%$, respectively. We translate all limits to $95\%$ using Poisson statistics. Our rate limit is consistent with the local BNS merger rate derived from gravitational wave observations by the LIGO-Virgo-KAGRA collaboration \citep[e.g.,][]{abbott_population_2023, abac_gwtc-40_2025}, resting a factor of 18 higher than their latest estimate. It also aligns with the rate implied by short GRB observations \citep[$160 - 352\,\mathrm{yr}^{-1}\,\mathrm{Gpc}^{-3}$; e.g.,][]{fong_decade_2015, dellavalle_gw170817_2018, zhang_peculiar_2018, dichiara_short_2020}.

It is important to emphasize that any derived upper limit is inherently model-dependent. For simplicity, and because it provides a robust empirical match to the only well-sampled event, our completeness analysis relies on the shock-cooling cocoon model of \citet{piro_evidence_2018}, which was calibrated to match the early, bright blue emission of SSS17a (AT~2017gfo). In the literature, a wide variety of models have been used to constrain rates. For instance, \citet{doctor_search_2017} used the Spectral Energy Distribution (SED) models of \citet{barnes_effect_2013} with varying velocity and mass parameters, while \citet{vanbemmel_optically_2025} employed the multi-component SSS17a (AT~2017gfo) best-fit model of \citet{villar_combined_2017}. Similarly, \citet{andreoni_constraining_2020} used the SSS17a (AT~2017gfo) best-fit model from \citet{dietrich_multimessenger_2020} alongside simple top-hat and linear decay models. Building on this, \citet{andreoni_fast-transient_2021} incorporated the viewing-angle-dependent models of \citet{bulla_possis_2019} and \citet{kasen_origin_2017}. More complex physical models are also available, incorporating dynamical ejecta, disk winds, magnetic fields, viscous evolution, and nuclear recombination \citep[e.g.,][]{metzger_time-dependent_2008, bauswein_systematics_2013, barnes_radioactivity_2016, dietrich_modeling_2017, siegel_three-dimensional_2017}. The observed KNe population is expected to be highly diverse, heavily dependent on the exact mass ratio, viewing angle, and ejecta composition of the merger \citep[e.g.,][]{metzger_electromagnetic_2010, barnes_effect_2013, barnes_radioactivity_2016}. If a significant fraction of KNe only have highly lanthanide-rich ejecta (and thus are strictly ``red,'' lacking the early blue optical peak) or are intrinsically fainter than SSS17a (AT~2017gfo), the ASAS-SN recovery efficiency would decrease, correspondingly raising the upper limit.

Because SSS17a (AT~2017gfo) is the only unambiguously confirmed, well-sampled KN associated with a GW event, the underlying luminosity function and color distribution of the KN population remain unconstrained. Our rate limit fundamentally assumes that the broader population shares similar peak optical properties to this single event and does not account for viewing angle differences in the early-time emission. Overcoming these uncertainties requires a large statistical sample; while high-cadence monitoring is ideal for fast transients, the immense depth and volume of next-generation facilities like the Vera Rubin Observatory are predicted to yield tens of un-triggered discoveries per year \citep[][]{perkins_searching_2026}. However, recognizing and classifying these faint detections will be a significant challenge for the community.

\section*{Acknowledgments}
DDD thanks Federica Chiti for helpful discussions. 

The Shappee group at the University of Hawai`i is supported with funds from NSF grant AST-2407205.
ASAS-SN is funded by Gordon and Betty Moore Foundation grants GBMF5490 and GBMF10501 and the Alfred P. Sloan Foundation grant G-202114192. 
CSK and KZS are supported by NSF grants AST-2307385 and AST-2407206. 
JFB is supported by NSF grant PHY-2310018. 
SD is supported by the National Natural Science Foundation of China (Grant No.\ 12133005). 
Parts of this research were supported by the Australian Research Council Centre of Excellence for Gravitational Wave Discovery (OzGrav), through project number CE230100016.
WBH acknowledges support from the National Science Foundation Graduate Research Fellowship Program under grant Nos. 2236415 and 1842402. 
Any opinions, findings, conclusions, or recommendations expressed in this material are those of the author(s) and do not necessarily reflect the views of the National Science Foundation.

\bibliography{zotero}
\bibliographystyle{mnras}

\appendix
\counterwithin{figure}{section}
\counterwithin{table}{section}
\renewcommand{\thefigure}{\thesection\arabic{figure}}
\renewcommand{\thetable}{\thesection\arabic{table}}
\renewcommand{\theHequation}{\thesection\arabic{equation}}
\renewcommand{\theHfigure}{\thesection\arabic{figure}}
\renewcommand{\theHtable}{\thesection\arabic{table}}

\section{Unclassified Transients Verification} \label{app:unclass_verify}

\begin{figure*}
    \centering
    \includegraphics[width=\textwidth]{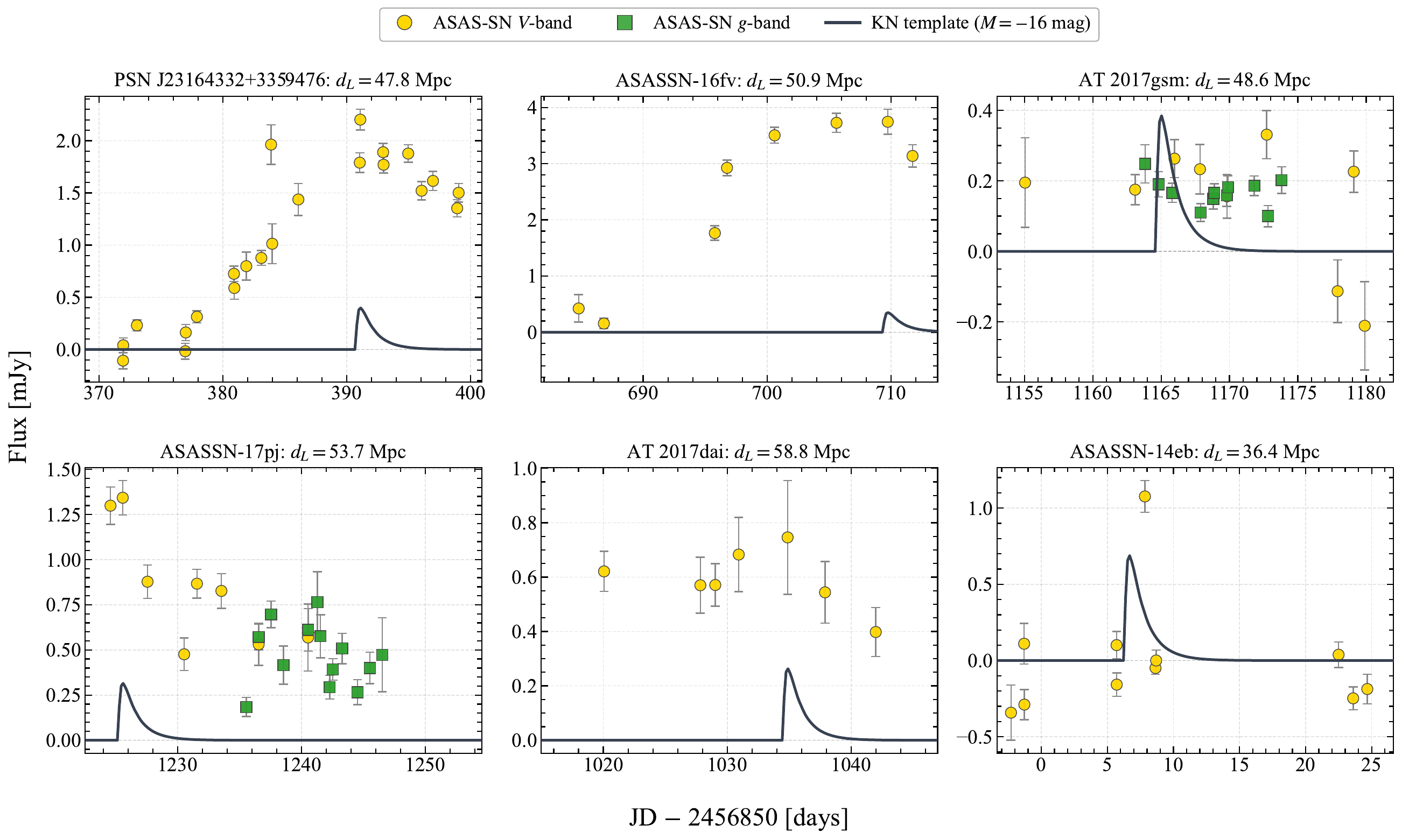}
    \caption{ASAS-SN light curves of the six unclassified transients rejected as KN candidates based on their photometric evolution. In the first five panels, the observed slow-evolving transients are compared to a nominal $M = -16.0$~mag KN template from \citet{piro_evidence_2018} (solid lines) placed at the distance of their respective host galaxies. The final panel shows ASASSN-14eb, a single-epoch detection. The solid line represents a forced KN template fit using the \citet{piro_evidence_2018} template.}
    \label{fig:unclass_lcs}
\end{figure*}

Our rate limits rely on the assumption that zero KNe were detected by ASAS-SN during the survey period. Given this, we first address why the only confirmed local KN associated with a GW event, SSS17a (AT~2017gfo), is absent from the ASAS-SN database. Based on an average $g$-band limiting magnitude of $18$~mag, SSS17a (AT~2017gfo) was intrinsically bright enough to be detected by ASAS-SN for approximately 1.5 days, and a single normal survey epoch can recover such an event out to $\sim60$~Mpc \citep{dejaeger_asas-sn_2022}. However, its non-recovery was entirely due to operational constraints. At the time of the merger in August 2017, only two of the five ASAS-SN network sites were operational, and the host galaxy's field was only observable early in the night. Consequently, the field was not scheduled for observation during that limited window, meaning SSS17a (AT~2017gfo) was never imaged by ASAS-SN.

We then conducted a systematic search within the historical ASAS-SN transient database to verify that no \textit{other} KN-like events were inadvertently ignored among the pool of unclassified transients.

We began with all sources formally flagged as transients by the ASAS-SN pipeline\footnote{\url{https://www.astronomy.ohio-state.edu/asassn/transients.html}}. From this initial list, we removed all spectroscopically confirmed objects (e.g., supernovae, tidal disruption events, active galactic nuclei, etc.) and high-confidence supernova candidates. To eliminate Galactic variables and stellar flares, we cross-matched the remaining unclassified sources against multi-wavelength stellar catalogs, including Gaia DR3 \citep{gaia_collaboration_gaia_2023}, SDSS DR16 \citep{ahumada_16th_2020}, WISE/AllWISE \citep{wright_wide-field_2010, mainzer_preliminary_2011}, DES DR2 \citep{abbott_dark_2021}, TESS \citep{paegert_tess_2021}, and GALEX \citep{bianchi_ultraviolet_2014}, removing any transient with a stellar counterpart within a $5\arcsec$ radius.

Next, we assessed the proximity of the remaining unclassified transients to known local galaxies. We cross-matched the transient coordinates with galaxies in the REGALADE catalog \citep[][]{tranin_catalog_2026}. Because ASAS-SN is not sensitive to SSS17a-like ``blue'' KNe beyond roughly $60~\mathrm{Mpc}$, we restricted our host galaxy search to this volume in REGALADE. To ensure we captured transients even in the remote outskirts of galaxies, we defined a spatial association as falling within two semi-major axes ($2R_{26}$) of the host, where $R_{26}$ corresponds to a surface brightness of $26~\mathrm{mag~arcsec^{-2}}$.

This stringent volume and host-association filtering left only eight unclassified candidates. Upon detailed inspection of their light curves and subtraction images, we ruled out all eight as viable KNe for the following reasons:
\begin{enumerate}[leftmargin=*]
    \item \textit{Slow Evolution (5 objects):} Five of the candidates exhibit photometric evolution that is significantly slower than the rapid timescales expected for KNe. Based on their light curve morphology, these are most likely unclassified supernovae. Their light curves, contrasted with expected KN evolutionary timescales, are shown in Figure~\ref{fig:unclass_lcs}.
    \item \textit{Morphological Incompatibility (1 object):} ASASSN-14eb is a single-epoch detection with no dithers. To test its viability, we forced a fit using the $V$-band KN light curve template. We fixed the absolute magnitude to $M = -16.0$~mag and set the distance to that of the associated host galaxy ($d_L = 36.4$~Mpc), with the time of peak as the only free parameter. The resulting fit is incompatible with the surrounding non-detections (see Figure~\ref{fig:unclass_lcs}), showing that the source is likely spurious.
    \item \textit{Image Subtraction Artifacts (1 object):} ASASSN-16pq was detected as a marginal $3\sigma$ source in only a single epoch. The signal was present in only one of the three ASAS-SN dithers comprising that epoch, indicating it is an image artifact, unusually shaped cosmic ray hit, or a short-duration optical glint caused by specular reflection from space debris.
    \item \textit{Low Significance / Unconfirmed (1 object):} ASASSN-19aea appeared as a single-epoch detection present in all three dithers. However, a 20-second $V$-band image taken with the UH88 telescope 23 hours post-discovery as part of the SCAT program \citep[][]{tucker_spectroscopic_2022} showed the source was $>1$~mag fainter than predicted by a KN-like decay. This is based on the KN light curve template fitting the peak 2.6~days prior to discovery. Additionally, it is located at the core of its host galaxy, and when evaluated against the local flux root-mean-square of the core, the detection significance drops to only $2\sigma$. 
\end{enumerate}

We therefore conclude that no viable KN candidates were missed among the unclassified ASAS-SN transients within $60~\mathrm{Mpc}$.

\end{document}